\documentclass[a4paper]{article}\usepackage{geometry}\usepackage{authblk}

\usepackage{ifthen}
\newboolean{isSciRep}

\usepackage{amsfonts}
\usepackage{amsmath}
\usepackage{color}
\usepackage{bbold}
\usepackage{hhline}
\usepackage[export]{adjustbox}
\usepackage{bbm}
\usepackage{makecell}
\usepackage{empheq}
\usepackage{cancel}

\newcommand{\epsi}{\varepsilon}
\newcommand{\mue}{\mu}

\newcommand{\Eth}{\vec{\mathcal{E}}}

\newcommand{\Hth}{\vec{\mathcal{H}}}

\newcommand{\Wsca}{W_\text{sca}}
\newcommand{\Wabs}{W_\text{abs}}
\newcommand{\Wext}{W_\text{ext}}

\newcommand{\Winc}{W_\text{inc}}

\newcommand{\Xext}{X_\text{ext}}
\newcommand{\Xsca}{X_\text{sca}}
\newcommand{\Xconv}{X_\text{conv}}
\newcommand{\Xinc}{X_\text{inc}}

\newcommand{\XX}{\mathbbm{X}}
\newcommand{\XSCA}{\XX_\text{sca}}
\newcommand{\XEXT}{\XX_\text{ext}}

\newcommand{\WW}{\mathbbm{W}}
\newcommand{\WSCA}{\WW_\text{sca}}
\newcommand{\WEXT}{\WW_\text{ext}}

\newcommand{\Ethinc}{\Eth_\text{inc}}
\newcommand{\Hthinc}{\Hth_\text{inc}}

\newcommand{\Ethsca}{\Eth_\text{sca}}
\newcommand{\Hthsca}{\Hth_\text{sca}}

\renewcommand{\vec}[1]{\mathbf{#1}}

\newcommand{\colvecTwo}[2]{
    \left( \begin{array}{c}
        #1 \\ #2
    \end{array} \right)}
\newcommand{\matrixTwo}[4]{
    \left( \begin{array}{cc}
        #1 & #2 \\ #3 & #4
    \end{array} \right)}

\newcommand{\xx}{\vec{x}}

\newcommand{\real}[1]{\operatorname{Re}\left(#1\right)}
\newcommand{\imag}[1]{\operatorname{Im}\left(#1\right)}

\newcommand{\norm}[1]{\left|#1\right|}
\newcommand{\normD}[1]{\left|\left|#1\right|\right|}

\renewcommand{\exp}[1]{e^{#1}}

\newcommand\blankfootnote[1]{%
  \let\thefootnote\relax\footnotetext{#1}%
  \let\thefootnote\svthefootnote%
}
\usepackage{hyperref}

\usepackage{pgfplotstable}
\usepackage{pgfplots}
\usepackage{tikz}
\usetikzlibrary{positioning,arrows}
\usetikzlibrary{calc,shapes}

\newcommand{\change}[1]{#1}
\newcommand{\changeTwo}[1]{#1}

\title{\change{Optical Chirality of Time-Harmonic Wavefields for Classification of Scatterers}}

\author[1,2,*]{Philipp Gutsche}
\author[3]{Manuel Nieto-Vesperinas}

\affil[1]{Freie Universit\"at Berlin, Mathematics Institute, Berlin, 14195, Germany}
\affil[2]{Zuse Institute Berlin, Computational Nano Optics, Berlin, 14195, Germany}
\affil[3]{Instituto de Ciencia de Materiales de Madrid,
	Consejo Superior de Investigaciones Cient\'ificas, Madrid, 28049, Spain}

\affil[*]{gutsche@zib.de}

\ifthenelse{\boolean{isSciRep}}{
}{}

\ifthenelse{\boolean{isSciRep}}{}{
\begin{document}\maketitle
\blankfootnote{\textit{Original Version:}
Gutsche, P. \& Nieto-Vesperinas, M. Optical Chirality of Time-Harmonic Wavefields for Classification of Scatterers.
Sci.~Rep.~\textbf{8}, 9416 (2018). \href{https://doi.org/10.1038/s41598-018-27496-w}{https://doi.org/10.1038/s41598-018-27496-w} }
}

\begin{abstract}
	\change{We derive expressions for the scattering, extinction and conversion of the chirality of monochromatic light scattered by bodies which are characterized by a $T$-matrix.}
	In analogy to the conditions obtained from the conservation of energy, these quantities enable the classification of arbitrary scattering objects due to their full,
	i.e.~either chiral or achiral, electromagnetic response. To this end, we put forward and determine the concepts of duality and breaking of duality symmetry,
	anti-duality, helicity variation, helicity annhiliation and the breaking of helicity annihilation.
	Different classes, such as chiral and dual scatterers, are illustrated in this analysis with model examples of spherical and non-spherical shape.
	As for spheres, these concepts are analysed by considering non-Rayleigh dipolar dielectric particles of high refractive index,
	which, having a strong magnetic response to the incident wavefield, offer an excellent laboratory to test and interpret such changes in the chirality of the illumination.
	In addition, comparisons with existing experimental data are made. 
\end{abstract}

\ifthenelse{\boolean{isSciRep}}{
	\begin{document}
	\flushbottom
	\maketitle
	\thispagestyle{empty}
}{}

\section*{Introduction}

Progress in designing spatially structured electromagnetic wavefields \cite{andrews2012} is giving rise to an increasing interest in electromagnetic waves
with twists of their polarization and wavefronts, i.e.~with spin and orbital angular momenta \cite{allen2003,allen1999,schaeferling2012}.
This complex shaped light is receiving substantial attention because of its potential for probing light-matter interactions, with additional information contents
like in new circular dichroism (CD) techniques in chiroptical spectroscopy \cite{schellman1975, richardson1977}
and spin-orbit phenomena \cite{vuong2010,bliokh2015,sukhov2015,hakobyan2015}, including Hall effects \cite{hosten2008,bliokh2015b}.
Additionally, such fields with angular momentum are of potential importance as communication vehicles with larger numbers of degrees of freedom \cite{andrews2013,osullivan2012,krenn2016}.

Considering light as a probe, the relationship between its chirality and that of matter is of great importance in the understanding of their mutual interactions \cite{bradshaw2015,nieto2017}.
Related magnetoelectric effects associated with the breaking of dual, P- and T-symmetries have been studied \cite{bliokh2014}.
However, procedures that enhance energy transfer (e.g.~F\"orster resonance energy transfer, FRET) between molecules \cite{vincent2011} and CD measurements are often hindered by very weak signals.
\change{
The sensitivity of such experiments is enhanced by increasing the helicity of the illuminating field \cite{tang2010, tang2011}, using either particles with plasmon resonances \cite{guzatov2012,alaeian2015,hentschel2017,kramer2017} or magneto-dielectric particles \cite{garcia2013}, or by means of near-field hot spots between plasmonic
nanoparticles on tailoring the incident polarization \cite{wang2015}.
}
Further strategies for strong chiroptical effects include thermally-controlled chirality in hybrid THz metamaterials \cite{lv2016}
and enhancing the interference of electric and magnetic dipoles of the excited molecule \cite{hu2016}.

In recent work, we established consequences of the continuity equation that governs the conservation of electromagnetic chirality of light
and other electromagnetic waves \cite{lipkin1964,bliokh2011,cameron2012}. In this way, we introduced an optical theorem which describes the extinction of helicity of
time-harmonic wavefields on scattering and absorption by arbitrary scatterers \cite{nieto2015} and that shows the connections between the chirality of the illuminating wave
and that of the scattering object. This yields a unified formulation of CD \cite{nieto2017,nieto2017a} and enables the introduction of a chirality enhancement factor \cite{nieto2017a,gutsche2017}
which is the chiral analogue of the Purcell factor for the emitted energy from nanostructures in inhomogeneous environments. In addition, the continuity equation of chirality conservation
was extended to twisted fields near nanostructures, as well as to arbitrary configurations \cite{gutsche2016,poulikakos2016}.
\change{We emphasize that the optical chirality of monochromatic, i.e.~time-harmonic, fields addressed in this work is equivalent to their helicity \cite{cameron2012,nieto2017}.
Both scalars differ only by a factor which is the square of the wavenumber \cite{nieto2015}. Since the former quantity has been extensively used in the literature after Ref.~\cite{tang2010},
we employ both terms interchangeably.}

\change{In this paper, we start from these concepts to classify the optical response of nanoparticles to monochromatic wavefields.}
These isolated scatterers are well described by their $T$-matrix \cite{mishchenko2002}.
In analogy, and complementary, to the $T$-matrix based standard classification of scatterers into lossless and lossy particles \cite{ru2013}, we now propose an analysis stemming
from the variation of helicity of chiral illuminating fields, which characterizes the major changes in the helicity of the incident wave with respect to that of the scattered field.
To this end, we first establish the chiral quantities for this general formulation, and develop a matrix-based formalism for the illumination with light of well-defined helicity \cite{fernandez2014}.
\change{
Then we put forward scalars which provide insight into the variation of incident circular polarization, including breakings of duality, anti-duality, annihilation of helicity,
as well as the helicity variation.
These concepts allow us to establish the following three classes of scattering bodies: helicity-keeping and helicity-flipping scatterers,
which are weaker forms of dual and anti-dual bodies, respectively. The third class which we introduce is that of helicity-annihilating particles.
}

Our findings are first illustrated by numerical studies of spherical objects described by Mie theory. We thus start by addressing a magneto-dielectric, dipolar in the broad sense \cite{nieto2017a}
(i.e. non-Rayleigh) silicon nanoparticle, as well as an isotropic chiral one built from  it. As we shall show, the existence of a strongly induced magnetic dipole,
in addition to the usual electric one, confers unique characteristics  to this kind of scatterer as regards the rich variety of effects that it causes in the incident field chirality.
Phenomena that, in turn, constitute signatures of the body magnetodielectric properties. Finally, the chiroptical behaviour of anisotropic particles is discussed with these novel quantities,
studying both an achiral ellipsoidal-shaped body and a gold nanoparticle  which was  experimentally investigated before \cite{mcpeak2014}.

\section*{Results}

\subsection*{$T$-Matrix Formalism}

The electromagnetic response of a scattering object to an external illumination in an extinction experiment,
i.e.~scattering plus absorption or conversion,
can be described by a matrix $T$ which gives the relation of the incident and scattered fields, $\vec{i}$
and $\vec{s}$ respectively\cite{mishchenko2002}: $T \cdot \vec{i} = \vec{s}$.

For isolated scatterers, vector spherical harmonics (VSHs) $\vec{M}^{(l)}$ and $\vec{N}^{(l)}$ ($l=1,3$) are a suitable
basis. VSHs are rigorous solutions of Maxwell's equations and there are two distinct classes.
In one class are the incident fields ($\vec{p}, \vec{q}$), the other class ($\vec{a}, \vec{b}$) pertains to the scattered fields
which obey the radiation condition.
\change{
Accordingly, the incident electric $\Ethinc$ and magnetic $\Hthinc$ time-harmonic fields are given by \cite[Sec.~9.7]{jackson1998}
\begin{align}
	\Ethinc(\xx,t) = \exp{-i\omega t} \sum_{m=1}^{\infty}\sum_{n=-m}^{n=m} p_{mn} \vec{N}_{mn}^{(1)}(\xx) + q_{mn} \vec{M}_{mn}^{(1)}(\xx)
  ,~~~
	\Hthinc(\xx,t) = -\frac{i}{Z} \exp{-i\omega t}  \sum_{m=1}^{\infty}\sum_{n=-m}^{n=m} p_{mn} \vec{M}_{mn}^{(1)}(\xx) + q_{mn} \vec{N}_{mn}^{(1)}(\xx)
		\label{eq:inc}
\end{align}
with the wave impedance $Z = \sqrt{\mue_0 \mue/(\epsi_0 \epsi)}$. While the scattered fields are
\begin{align}
  \Ethsca(\xx,t) = \exp{-i \omega t} \sum_{m=1}^{\infty}\sum_{n=-m}^{n=m} a_{mn} \vec{N}_{mn}^{(3)}(\xx) + b_{mn} \vec{M}_{mn}^{(3)}(\xx)
  ,~~~
	\Hthsca(\xx,t) = -\frac{i}{Z} \exp{-i\omega t} \sum_{m=1}^{\infty}\sum_{n=-m}^{n=m}  a_{mn} \vec{M}_{mn}^{(3)}(\xx) + b_{mn} \vec{N}_{mn}^{(3)}(\xx).
		\label{eq:sct}
\end{align}
The index $m$ indicates the multipole order and it is associated to the eigenvalue $m(m+1)$ of the squared orbital angular momentum operator $L^2$ in the spherical harmonic basis
$Y_{mn}(\theta,\phi)$ \cite[Sec.~9.7]{jackson1998} ($\theta$ and $\phi$ being the polar and azimuthal angles of the position vector $\vec{x}$).
The series above will be truncated at an order $M$ in the following calculations.
The index $|n|<m$ is related to the azimuthal behaviour of the VSHs. The VSHs $\vec{N}_{mn}^{(3)}$ and $\vec{M}_{mn}^{(3)}$ are proportional to the electric and magnetic
outgoing multipolar fields, respectively.  On the other hand, the VSHs $\vec{N}_{mn}^{(1)}$ and $\vec{M}_{mn}^{(1)}$ of the source-free incident wave are proportional
to the electric and magnetic multipolar fields with both outgoing and incoming components \cite{nieto2006}.
Because of this, \eqref{eq:inc} and \eqref{eq:sct} represent the incident and scattered fields in the \textit{parity basis}, namely that of eigenmodes of electric and magnetic nature.}  

In matrix notation, the coefficients of the series expansions \eqref{eq:inc} and \eqref{eq:sct} are related to each other through the $T$-matrix \cite{waterman1968,mishchenko2002}:
\begin{align}
  T \colvecTwo{\vec{p}}{\vec{q}} =
  \matrixTwo{T_{ee}}{T_{me}}{T_{em}}{T_{mm}} \colvecTwo{\vec{p}}{\vec{q}} &= \colvecTwo{\vec{a}}{\vec{b}},
	\label{eq:T}
\end{align}
where $T$ has been subdivided into electric $T_{ee}$, magnetic $T_{mm}$ and cross electric-magnetic $T_{em},T_{me}$ matrices.

The scattering solution for isotropic spherical particles with relative permittivity $\epsi$ and relative permeability $\mue$
are given analytically \cite{bohren1940}.
\change{
Furthermore, if the sphere is optically active, the refractive indices differ for right and left
circularly polarized illumination, being $n_R$ and $n_L$, respectively. Using the achiral refractive index $n = \sqrt{\epsi\mue}$ and
assuming that particle is reciprocal,
its optical behaviour is described with the Pasteur, or chirality, parameter $\kappa \in [-1,1]$ \cite[Eq.~(2.85)]{lindell1994} as
$n_R = n ( 1 + \kappa )$ and $n_L = n ( 1 - \kappa )$.
}

For a geometrically isotropic sphere, the respective submatrices of the $T$-matrix in \eqref{eq:T} are diagonal.
In the case of reciprocal materials, we have additionally $T_{em} = T_{me}$.
The main diagonal elements of $T_{ee}$, $T_{mm}$ and $T_{em}$ are given in the Methods section.

\subsubsection*{Energy Conservation}

Assuming the scatterer is embedded in a lossless medium, the conservation of energy
predicts a scattered $\Wsca$, extinction $\Wext$ and absorption $\Wabs$ of energy related by
$\Wext = \Wsca + \Wabs$.
\change{In the VSH basis, these quantitites are reduced to the following expressions \cite[Eq.~(5.18a,5.18b)]{mishchenko2002}.
Note that our notation for the role of $(\vec{p},\vec{q})$ and $(\vec{a},\vec{b})$ is interchanged with respect to that of Ref.~\cite{mishchenko2002}, using instead that of Ref.~\cite{fruhnert2017}.
For the sake of brevity, we henceforth omit the variation ranges of the indices $m$ and $n$ in the series representations and obtain}
\begin{align}
  \Wsca &= \frac{1}{2 k^2 Z} \sum_{mn} |a_{mn}|^2 + |b_{mn}|^2 = \frac{1}{2 k^2 Z} \left( |\vec{a}|^2 + |\vec{b}|^2 \right)
	, \label{eq:Wsca} \\
  \Wext &= -\frac{1}{2 k^2 Z} \sum_{mn} \real{ p^*_{mn} a_{mn} + q^*_{mn} b_{mn} }
    = \frac{1}{2 k^2 Z} \frac{1}{2} \left( \vec{p}^H \vec{a} + \vec{q}^H \vec{b} + \vec{a}^H \vec{p} + \vec{b}^H \vec{q} \right),
		\label{eq:Wext}
\end{align}
where $^H$ denotes the Hermitian adjoint, i.e.~matrix transposition and complex conjugation,
and $k=n_s \omega/c_0$ is the wavenumber, with $n_s$ being the refractive index of the embedding medium.
\change{
Since we use normalized VSHs, here and in the following we drop the terms $|\vec{N}_{mn}|^2 = |\vec{M}_{mn}|^2 = 1 (V/m)^2$.
Taking this unity factor and its dimension into account, the unit of the integrated energy fluxes $\Wsca$ and $\Wext$ is watt, as expected.
}
Eqs.~\eqref{eq:Wsca} and \eqref{eq:Wext} may be written as  
\begin{align}
  2 k^2 Z \Wsca &=
    \left(\vec{p}^H \vec{q}^H\right)
      T^H T
    \colvecTwo{\vec{p}}{\vec{q}}
    =
    \left(\vec{p}^H \vec{q}^H\right)
      \WSCA
    \colvecTwo{\vec{p}}{\vec{q}}
  , \label{eq:Wscaa} \\  
  2 k^2 Z \Wext &=
    -\left(\vec{p}^H \vec{q}^H\right)
      \frac{1}{2} (T^H + T)
    \colvecTwo{\vec{p}}{\vec{q}}
    =
    \left(\vec{p}^H \vec{q}^H\right)
      \WEXT
    \colvecTwo{\vec{p}}{\vec{q}}. \label{eq:Wexta}
\end{align}
The quadratic matrices $\WSCA$ and $\WEXT$ have dimension $2M(M+2)$ for multipole orders $m=1,...,M$.
From the conservation of energy, the following conditions apply \cite[Eq.~(10)]{ru2013}
\begin{align}
  \text{Lossless scatterer: } &\WEXT = \WSCA, \\
  \text{Lossy scatterer: } &\left(\WEXT - \WSCA\right) \text{ is HPD},
\end{align}
where HPD means Hermitian positive definite.
Note that both $\WSCA$ and $\WEXT$ are Hermitian by construction.
\change{
These general matrices provide the illumination-independent information on scattering and extinction of energy by an isolated scatterer whose optical response is described by its $T$-matrix.
Eq.~\eqref{eq:Wscaa} indicates that multiplying $\WSCA$ on the left and right with the vector of coefficients $(\vec{p}, \vec{q})$ of the incident light gives the scalar scattered energy $\Wsca$,
for the specific illumination with $(\vec{p}, \vec{q})$. The same holds for the matrix $\WEXT$ and the energy $\Wext$ extinguished from the incident field [cf.~Eq.~\eqref{eq:Wexta}]
as well as for the matrices $\XSCA$ and $\XEXT$, which represent the scattered and extinguished optical chirality and are introduced in \eqref{eq:XXsca} and  \eqref{eq:XXext} below.
}

\subsubsection*{Chirality Conservation}

\change{
The conservation law of optical chirality \cite{gutsche2016,nieto2015}
states that the scattered chirality $\Xsca$ and the extinguished chirality (or chirality extinction) $\Xext$, as well as
the converted chirality (or chirality conversion) $\Xconv$ are related by
$\Xext = \Xsca + \Xconv$.
}
Note that as shown in Ref.~\cite{gutsche2016} $\Xconv$ is a field chirality that may be either absorbed or
generated on scattering of the incident wave by the body. This is why it is named \textbf{chirality conversion}. 

\change{
The optical chirality density in the near-field is proportional to the excitation rate of chiral molecules \cite{tang2010}.
Here, we study the integrated optical chirality flux density yielding extinction $\Xext$ and scattered $\Xsca$ chirality, as well as its conversion $\Xconv$.
The scattered optical chirality flux \textit{density} is proportional to the difference of the circular polarization components of the scattered field
at a specific point in space, thus being the third Stokes parameter. 
Its \textit{integral}, the scattered chirality $\Xsca$, is the angular average of the differential circular polarization of scattering.
That is, $\Xsca=0$ both for locally achiral light (e.g.~linearly polarized plane waves), as well as for light sources which emit equal parts of right and left
circular polarization in different directions (e.g.~circularly polarized electric dipoles).
\\ \indent
In order to classify the chiroptical response of isolated scatterers, we henceforth establish these quantities in the VSH basis:
}
\begin{align}
  \Xsca &= \frac{1}{2 k Z} \sum_{mn} 2 \real{ a_{mn}^* b_{mn} }
    = \frac{1}{2 k Z} \left( \vec{a}^H \vec{b} + \vec{b}^H \vec{a} \right),
		\label{eq:Xsca}
	\\
  \Xext &= -\frac{1}{2 k Z} \sum_{mn} \real{ p^*_{mn} b_{mn} + q^*_{mn} a_{mn} }
    = -\frac{1}{2 k Z} \frac{1}{2} \left( \vec{p}^H \vec{b} + \vec{q}^H \vec{a} + \vec{b}^H \vec{p} + \vec{a}^H \vec{q} \right).
\end{align}
And they may also be written as
\begin{align}
  2 k Z \Xsca &=
    \left(\vec{p}^H \vec{q}^H\right)
      \matrixTwo{ T_{ee}^H T_{em} + T_{em}^H T_{ee} }{ T_{ee}^H T_{mm} + T_{em}^H T_{me} }{
        T_{mm}^H T_{ee} + T_{me}^H T_{em} }{ T_{me}^H T_{mm} + T_{mm}^H T_{me} }
    \colvecTwo{\vec{p}}{\vec{q}}
    =
    \left(\vec{p}^H \vec{q}^H\right)
      \XSCA
    \colvecTwo{\vec{p}}{\vec{q}}
  	\label{eq:XXsca}
	, \\
  2 k Z \Xext &=
    -\left(\vec{p}^H \vec{q}^H\right)
      \frac{1}{2} \matrixTwo{ T_{em}+T_{em}^H }{ T_{ee}^H+T_{mm} }{ T_{ee}+T_{mm}^H }{ T_{me}+T_{me}^H }
    \colvecTwo{\vec{p}}{\vec{q}}
    =
    \left(\vec{p}^H \vec{q}^H\right)
      \XEXT
    \colvecTwo{\vec{p}}{\vec{q}}.
  	\label{eq:XXext}
\end{align}

\subsubsection*{Incident Light of Well-Defined Helicity}

\change{
Of particular interest are wavefields of well-defined helicity \cite{fernandez2013}, i.e.~those whose plane wave components \cite{bliokh2011,nieto2017a}
all have the same helicity handedness of circular polarization with respect to their wavevector. In the following, we discuss the helicity of the incident light
which is given by the VSH coefficients $(\vec{p},\vec{q})$ in \eqref{eq:inc}.
\\ \indent
The transformation \cite{fernandez2014,zambrana2016,fernandez2018}:
\begin{align}
	\vec{A}_{nm}^{(1\,\pm)}(\xx)=\frac{\vec{N}_{mn}^{(1)}(\xx) \pm \vec{M}_{mn}^{(1)}(\xx)}{\sqrt{2}},
		\label{eq:helitrans}
\end{align}
changes the representation \eqref{eq:inc} of the incident wavefield in the parity basis to the \textit{helicity basis}
in which they have positive [sign $+$ in \eqref{eq:helitrans}] or negative [sign $-$ in \eqref{eq:helitrans}] \textit{well-defined helicity}.
\\ \indent
In this framework, the conditions for fields of well-defined helicity may be straightforwardly derived from the relation of the integrated energy 
and chirality introduced in the previous section. The incident energy is $2 k^2 Z \Winc = |\vec{p}|^2 + |\vec{q}|^2$ [cf.~\eqref{eq:Wsca}
replacing $(\vec{a},\vec{b})$ by $(\vec{p},\vec{q})$], whereas the incident chirality is $2 k Z \Xinc = 2 \real{\vec{p}^* \cdot \vec{q}}$
[cf.~\eqref{eq:Xsca}].
\\ \indent
For incident fields of well-defined helicity, it is required that $\Xinc = \pm k \Winc$, where as in \eqref{eq:helitrans} the sign $\pm$ denotes a state
of positive and negative helicity, respectively. Accordingly, a positive helicity state requires that $\vec{q} = \vec{p}$ and a
negative helicity state is characterized by $\vec{q} = -\vec{p}$. Notice that both conditions are indepedent of each other, so that the coefficients of
a positive helicity state are not related to those of a negative helicity state. That is why we denote $\vec{p}^+$ and $\vec{p}^-$ the coefficients
of an arbitrary field of well-defined positive and negative helicity, respectively.
Summarizing, it holds either $\vec{q}^+ = \vec{p}^+$ or $\vec{q}^- = -\vec{p}^-$ for incident light of well-defined positive or negative helicity, respectively.
}

\change{
Therefore here and throughout the entire paper, the superscripts $\pm$ denote illumination with light of well-defined helicity.
In what follows, we shall analyze the response of an arbitrary scatterer to incident light of either positive or negative helicity.
Specifically, we investigate the scattered energy $\Wsca^+$ or $\Wsca^-$, as well as the scattered chirality $\Xsca^\pm$, for illuminating light of positive or negative helicity, respectively.
}

For $2 k^2 Z \Wsca^\pm = ({\vec{p}^\pm})^H \WSCA^\pm \vec{p}^\pm$
and $2 k Z \Xsca^\pm = ({\vec{p}^\pm})^H \XSCA^\pm \vec{p}^\pm$, one has that
\newcommand{\highl}[1]{\textcolor{black}{#1}}
\begin{align}
  \WSCA^\pm &= \highl{+} \left\{ \left(T_{\highl{e}}^\pm\right)^H T_e^\pm + \left(T_{\highl{m}}^\pm\right)^H T_m^\pm \right\}
		\label{eq:WSCApm} , \\
  \XSCA^\pm &= \highl{\pm} \left\{ \left(T_{\highl{m}}^\pm\right)^H T_e^\pm + \left(T_{\highl{e}}^\pm\right)^H T_m^\pm \right\}
		\label{eq:XSCApm}, 
\end{align}
where $T_e^\pm = T_{ee} \pm T_{me}$ and $T_m^\pm = T_{mm} \pm T_{em}$.
For $2 k^2 Z \Wext^\pm = ({\vec{p}^\pm})^H \WEXT^\pm \vec{p}^\pm$
and $2 k Z \Xext^\pm = ({\vec{p}^\pm})^H \XEXT^\pm \vec{p}^\pm$, it holds
\begin{align}
  \WEXT^\pm
    = \pm\XEXT^\pm
    = -\frac{1}{2} \left\{ T_e^\pm + \left(T_e^\pm\right)^H + T_m^\pm + \left(T_m^\pm\right)^H \right\}.
  \label{eq:extHelicityMatrices}
\end{align}
For randomly oriented isolated scatterers, averaging over all illumination directions is of interest. This is because in dilute solutions,
where multiple scattering can be neglected, the experimental results are dominated by the averaged response of a single particle \cite{mcpeak2014}.
Let $\Wext^\pm(\theta,\phi)$ be the energy extinction of an incident circularly ($\pm$)-polarized (CPL) plane wave with propagation direction given by $\theta$ and $\phi$.
From the expansion of plane waves into vector spherical harmonics \cite[Eq.~(C57)]{mishchenko2002}, one derives that the averaged energy extinction is given by
\begin{align}
  \overline{\Wext^\pm} = \frac{1}{2\pi^2} \int \int \Wext^\pm(\theta,\phi) d\theta d\phi
    = \frac{\pi}{k^2 Z} \sum_{ii} \left(\WEXT^\pm\right)_{ii},
  \label{eq:avgWext}
\end{align}
where $\left(\WEXT^\pm\right)_{ii}$ are the diagonal elements of the extinction energy matrices of well-defined helicity states \eqref{eq:extHelicityMatrices}.
Similarly, it follows
\begin{align}
  \overline{\Wsca^\pm} = \frac{\pi}{k^2 Z} \sum_{ii} \left(\WSCA^\pm\right)_{ii}
	,~~~
  \overline{\Xsca^\pm} = \frac{\pi}{k Z} \sum_{ii} \left(\XSCA^\pm\right)_{ii}
  ,~~~
	\overline{\Xext^\pm} = \frac{\pi}{k Z} \sum_{ii} \left(\XEXT^\pm\right)_{ii}.
		\label{eq:avgXext}
\end{align}

\subsection*{Definitions for Classification of Scatterers}

\subsubsection*{Duality and Anti-Duality}

The scattered field of a \textbf{dual} scatterer has the same (well-defined) helicity as the incident field, i.e.~$\Xsca^+ = k \Wsca^+$ for
positive incident helicity, and $\Xsca^- = -k \Wsca^-$ for negative incident helicity. In matrix notation, it follows
$
  \XSCA^\pm \mp \WSCA^\pm
  =
  \left(T_e^\pm - T_m^\pm\right)^H
  \left(T_e^\pm - T_m^\pm\right)
  = 0
$.
In general, we establish the {\bf duality breaking} $\cancel{d} \in [0,1]$ of the scattering object
with \changeTwo{an} arbitrary matrix norm $\normD{\cdot}$ as
\begin{align}
  \cancel{d} = \max_\pm \left( \frac{\normD{T_e^\pm - T_m^\pm}}{\normD{T_e^\pm} + \normD{T_m^\pm}} \right).
	\label{eq:dualBreak}
\end{align}
\change{
Our formalism is independent of the choice of the specific
norm. Due to its numerical robustness, we choose the 2-norm $\normD{A}_2$ of
a matrix $A$ which is given by the largest singular value of $A$.
}
Note in this connection that if the Frobenius norm were chosen in Eq.~\eqref{eq:dualBreak}, $\cancel{d}$ would be
similar to the definition established in
\cite[Eq.~(23)]{fruhnert2017},
\cite[Eq.~(32)]{fernandez2016} and
\cite[Eq.~(2)]{fernandez2015}.

On the other hand, if the scattered field has the opposite (well-defined) helicity with respect to that of the incident field,
i.e.~$\Xsca^+ = -k \Wsca^+$ for
positive incident helicity and $\Xsca^- = k \Wsca^-$ for negative incident helicity, we shall say that the scatterer is  \textbf{anti-dual}.
In matrix notation, it follows
$
  \XSCA^\pm \pm \WSCA^\pm
  =
  \left(T_e^\pm + T_m^\pm\right)^H
  \left(T_e^\pm + T_m^\pm\right)
  = 0
$.
Thus, in general, we herewith put forward the \textbf{anti-duality breaking} $\cancel{a} \in [0,1]$ as
\begin{align}
  \cancel{a} = \max_\pm \left( \frac{\normD{T_e^\pm + T_m^\pm}}{\normD{T_e^\pm} + \normD{T_m^\pm}} \right).
\end{align}

\subsubsection*{Helicity Variation}
Given an incident field of well-defined helicity, in general the helicity of the scattered field may have the same sign as the incident one
but may not be well-defined, i.e.~its angular spectrum of plane wave components\cite{bliokh2011,nieto2017a} is composed of some
with positive and some with negative helicity.
If $\Xsca$ has the same sign as that of the incident field (e.g.~$\Xsca^+ > 0$ for positive incident helicity),
we put forward the term \textbf{helicity-keeping} classifying such scatterers.
By contrast, a \textbf{helicity-flipping} scatterer (e.g.~$\Xsca^+ < 0$ for positive incident helicity) changes the sign of the incident helicity to its opposite value. If an incident chiral field is scattered
into a purely achiral field with $\Xsca^\pm=0$, we shall call the scatterer \textbf{helicity-annihilating}.
Therefore, we establish the following classification of scattering bodies:
\begin{align}
  \text{Perfectly Helicity-Keeping: } &
    \XSCA^+ \text{ (HPD) and  }
    \XSCA^- \text{ (HND)},
  \\
  \text{Perfectly Helicity-Flipping: } &
    \XSCA^+ \text{ (HND) and  }
    \XSCA^- \text{ (HPD)},
  \\
  \text{Perfectly Helicity-Annihilating: } &
    \XSCA^\pm = 0.
\end{align}
We recall that HPD and HND stand for Hermitian positive and negative definite matrix, respectively.
From \eqref{eq:WSCApm}, \eqref{eq:XSCApm} and the triangle inequality, it follows that
$\normD{\WSCA^\pm} \leq \normD{T_e^\pm}^2 + \normD{T_m^\pm}^2$
and
$\normD{\XSCA^\pm} \leq 2 \normD{T_e^\pm} \normD{T_m^\pm}$.
Based on the former relations we establish the \textbf{breaking of helicity annhiliation} $\cancel{h}_a$ relative to the scattered energy:
\begin{align}
  \cancel{h}_a = \max_\pm \left( \frac{\normD{\XSCA^\pm}}{\normD{\WSCA^\pm}} \right).
\end{align}
This novel quantity vanishes for  linearly polarized scattered light, i.e.~$\cancel{h}_a=0$, as then scattering annihilates all incident helicity.
In contrast, the breaking of helicity annihilation is one for scattered light of well-defined helicity;
however, it is independent of the incident helicity, which means $\cancel{h}_a=1$ if the scattered light
has the same helicity as, or opposite to, the helicity as the incident light.

Further, we introduce the \textbf{helicity variation} $h_v \in [-1,1]$ by the eigenvalues $\lambda^{\XSCA^\pm}_i$ of
the chirality scattering matrix, namely, $\XSCA^\pm \vec{v}_i = \lambda^{\XSCA^\pm}_i \vec{v}_i$ as
\begin{align}
  h_v = \frac{1}{2} \left(
      \frac{\sum_i \lambda^{\XSCA^+}_i}{\sum_i \left|\lambda^{\XSCA^+}_i\right|}
      -
      \frac{\sum_i \lambda^{\XSCA^-}_i}{\sum_i \left|\lambda^{\XSCA^-}_i\right|}
    \right).
		\label{eq:heli_var}
\end{align}
\change{
According to this definition, a helicity-keeping scatterer has $h_v>0$, while a helicity-flipping one has $h_v<0$.
Note that a necessary condition for a  scatterer to be anti-dual is to be perfectly helicity-flipping, i.e.~$h_v=-1$. Conversely, for a dual scatterer the condition $h_v=1$
(\changeTwo{namely, to be perfectly} helicity-keeping)
is necessary. However, these two last conditions for $h_v$ are not sufficient for either duality or anti-duality, and thus they may be regarded as weaker forms of
these two latter properties.
}

Specifically, the helicity variation only takes the sign of the eigenvalues of $\XSCA$ into account.
For a dual scatterer not only the sign of the incident helicity has to be preserved, but the eigenvalues of $\XSCA$ and $\WSCA$
must be of equal absolute value. So, $h_v=1$ is a weaker condition than $\cancel{d}=0$ since  $h_v=1$ means that the scattered light
is dominated by the incident helicity, but it may not  possess well-defined helicity.
On the other hand, $\cancel{d}=0$ implies that the scattered light has well-defined helicity equal to that of the incident light.

\changeTwo{
As described above, perfectly helicity-keeping ($h_v=1$) and perfectly helicity-flipping ($h_v=-1$) are weaker conditions of duality and anti-duality.
These conditions may be further weakened yielding helicity-keeping ($h_v>0$) and helicity-flipping ($h_v<0$) objects.
Since the helicity variation \eqref{eq:heli_var} is a weighted average over all eigenvalues of $\XSCA$, $h_v$ describes the mean alteration of the sign of the scattered chirality for all possible
incident fields of well-defined helicity. For a helicity-keeping scatterer, the chirality scattering is dominated by incident fields for which the incident and the scattered
helicity equal one-another. It does not imply that the helicity of all possible incident fields is unchanged, however, the major contribution to scattering is due
to incident fields with preserved helicity.
Accordingly, for strongly scattering illuminations of well-defined helicity, the scattered helicity changes sign predominantly for a helicity-flipping scatterer.
}

\subsubsection*{Chirality}

Starting from \eqref{eq:avgWext},
the chirality $c$ and the $g$-factor of a scatterer are defined as
\begin{align}
  c = \overline{\Wext^+} - \overline{\Wext^-}
	,~~~
  g = \frac{\overline{\Wext^+} - \overline{\Wext^-}}{\overline{\Wext^+} + \overline{\Wext^-}},
		\label{eq:g}
\end{align}
\change{where the average bars should be understood as written in \eqref{eq:avgWext} and \eqref{eq:avgXext}.
Hence, these quantitites represent the plane-wave averaged values derived from the full $T$-matrix, and are
the corresponding generalizations for wide sense  dipolar particles \cite{tang2010}}. We notice that during the writing
of this manuscript, a $T$-matrix formalism introducing quantities similar to those
of Eq.~\eqref{eq:g} has been developed in Ref.~\cite{suryadharma2018}. 

\section*{Discussion of Examples}

\ifthenelse{\boolean{isSciRep}}{}{\subsection*{Isotropic Scatterer}}
\ifthenelse{\boolean{isSciRep}}{}{\subsubsection*{Dipolar Isotropic Scatterer}}

Electromagnetic scattering by non-Rayleigh dipolar isotropic scattering objects is  described by their electric and magnetic  polarizabilities, $\alpha_e$  and $\alpha_m$,
and by their cross electric-magnetic ones, $\alpha_{em}$ and $\alpha_{me}$.
We assume reciprocal
scatterers with $\alpha_{em} = -\alpha_{me}$.
With $a_1$, $b_1$ and $c_1$ being the electric, magnetic and cross electric-magnetic first Mie coefficients, it is well-known that
$\alpha_e = \frac{6i\pi\epsi_0}{k^3} a_1$
, $\alpha_m = \frac{6i\pi\mue_0}{k^3} b_1$
and $\alpha_{me} = \frac{6i\pi\sqrt{\epsi_0\mue_0}}{k^3} c_1$.
In terms of these polarizabilities, one has for the quantitites introduced above:
\begin{align}
  \Wext^\pm &\propto \imag{\alpha_e + \alpha_m} \pm 2 \real{\alpha_{em}},
		\label{eq:dip_Wext}
  \\
  \Wsca^\pm &\propto 2 \norm{\alpha_{em}}^2 \mp 2 \imag{\alpha_{em}^* \left\{\alpha_e + \alpha_m\right\}}
    + \norm{\alpha_e}^2 + \norm{\alpha_m}^2,
		\label{eq:dip_Wsca}
  \\
  \Xsca^\pm &\propto -2 \imag{\alpha_{em}^* \left\{\alpha_e + \alpha_m\right\}}
    \pm 2 \real{\alpha_e^* \alpha_m} \pm 2 \norm{\alpha_{em}}^2,
		\label{eq:dip_Xsca}
  \\
  \cancel{d} &= \max_\pm \left( \frac{\left|\alpha_e - \alpha_m\right|}
    {\left|\alpha_e \pm \alpha_{em}\right| + \left|\alpha_m \pm \alpha_{em}\right|} \right),
		\label{eq:dip_d}
  \\
  \cancel{a} &= \max_\pm \left( \frac{\left|\alpha_e + \alpha_m \pm 2 \alpha_{em}\right|}
    {\left|\alpha_e \pm \alpha_{em}\right| + \left|\alpha_m \pm \alpha_{em}\right|} \right),
		\label{eq:dip_a}
  \\
  c &\propto \real{\alpha_{em}},
  \\
  g &= \frac{2 \real{\alpha_{em}}}{\imag{\alpha_e + \alpha_m}}.
		\label{eq:dip_g}
\end{align}
In general, the four experimental observables $\Wext^\pm$ and $\Wsca^\pm$ are not sufficient to determine
the real and imaginary parts of those three polarizabilities. Therefore, we propose the additional measurement
of the scattered chirality $\Xsca^\pm$. This together with Eqs.~\eqref{eq:dip_Wext}-\eqref{eq:dip_g} should enable
the full optical characterization of a dipolar scattering object.

The $g$-factor expressed by \eqref{eq:g} is well-known: it is the dissymmetry factor\cite{craig1984,bohren1940}
of circular dichroism (see e.g.~\cite[cf.~Eq.~(6)]{tang2010}).
Furthermore, the chirality \change{(or helicity)} $c$, being the differential extinction due to incident circularly polarized light, is proportional
to the real part of the cross electric-magnetic polarizability $\alpha_{em}$ \cite{tang2010}.
This measures the optical activity of the scatterer like e.g.~a chiral nanoparticle or molecule \cite{bohren1940}.

\change{
A dipolar body with $\alpha_e = \alpha_m$ (and as stated above, $\alpha_{em} =-\alpha_{me}$) guarantees duality symmetry \cite{nieto2015};
thus it presents a vanishing duality breaking parameter ($\cancel{d} = 0$).
This requirement coincides with the  first Kerker condition
\cite{kerker1983,gomez2011,nieto2011,geffrin2012,fu2013,person2013,staude2013,zhang2015,decker2016,kuznetsov2016}
\changeTwo{according} to which there is zero backscattered intensity under plane wave illumination.
}
\change{
In fact, it has been shown that a dual scatterer produces zero-backscattering 
and that duality may be regarded as a generalization of the first
Kerker condition \cite{zambrana2013}.
}
On the other hand, the second Kerker condition \cite{kerker1983} for achiral lossless objects
($\alpha_{em}=0$) is $\imag{\alpha_e}=\imag{\alpha_m}$ and $\real{\alpha_e} = -\real{\alpha_m}$, yields a minimum of the forward scattered intensity \cite{nieto2011}
and also of the anti-duality breaking $\cancel{a}$.
A scatterer with strictly vanishing $\cancel{a}$ has been recognized as behaving as anti-dual \cite{zambrana2013}.
Note that as a consequence of \eqref{eq:dip_a}, \textit{a chiral reciprocal non-Rayleigh dipolar particle, i.e.  with
$\alpha_{em}\neq0$, cannot be anti-dual}.

\begin{figure}[ht!]
	\centering
	\includegraphics[width=.95\textwidth]{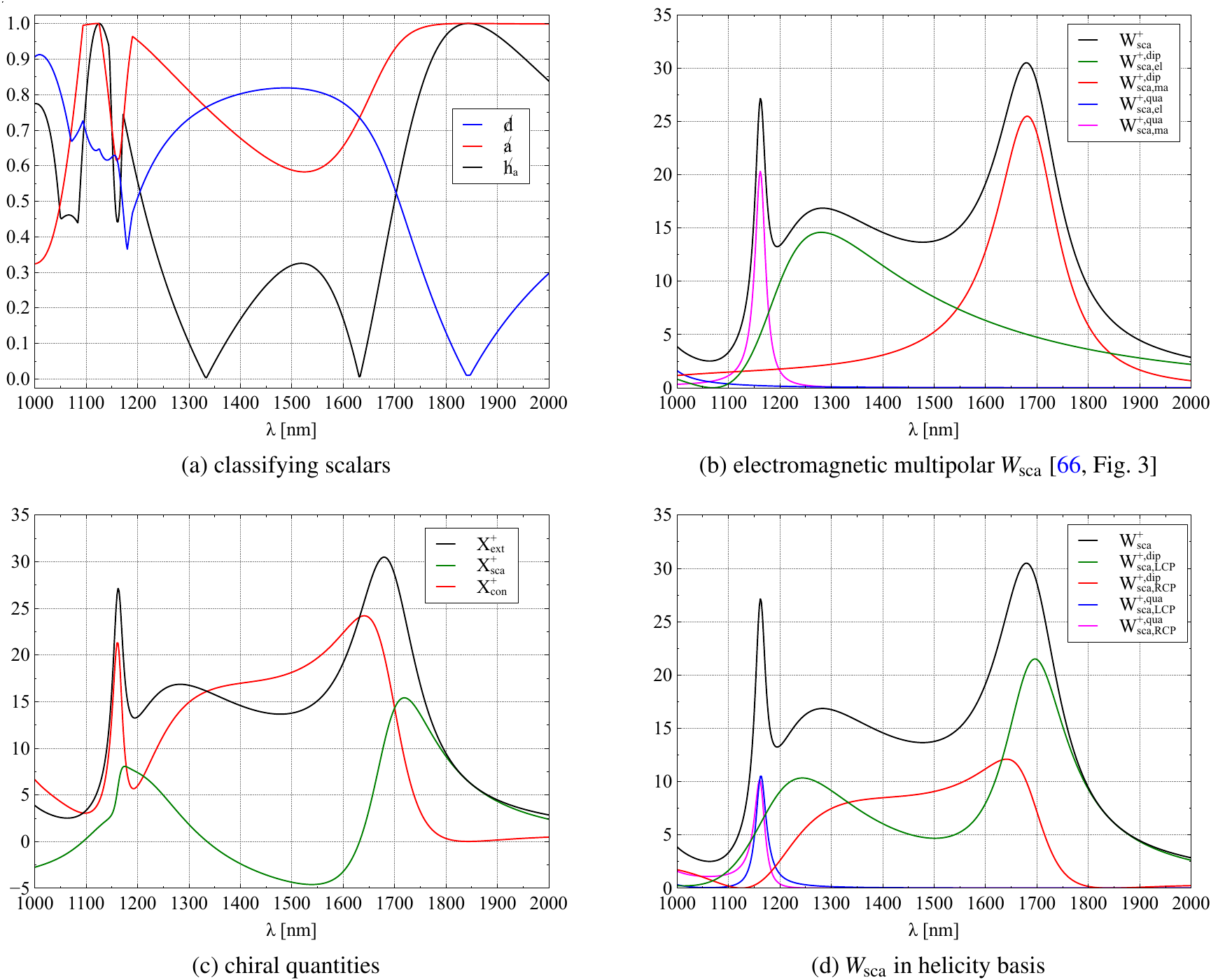}
	\caption{
		Spectra from a spherical Si particle of radius $r=230\text{nm}$ and refractive index $n=3.5$ illuminated by a left circularly polarized LCP (+) plane wave.
		(a) Breakings of duality $\cancel{d}$, anti-duality $\cancel{a}$ and helicity annihilation $\cancel{h}_a$. 
		(b) Scattered energies due to the excitation of each of the induced electric and magnetic dipoles (at 1280nm and 1680nm, respectively),
		and magnetic quadrupole (at 1160nm), and total scattered energy due to the superposition of their respective scattered fields (after \cite{garcia2011}).
		(c) Extinction, scattered, and conversion chiralities.
		(d) Multipolar scattered energies in the helicity basis (cf.~main text). \change{These are: due to the excitation of the RCP(-) dipole and quadrupole,
		$W^\text{+,dip}_\text{sca,RCP}$ (red line) and $W^\text{+,qua}_\text{sca,RCP}$ (pink line), respectively; as well as  due to the excitation of the
		LCP(+)  dipole and quadrupole, \change{$W^\text{+,dip}_\text{sca,LCP}$ (green line) and $W^\text{+,qua}_\text{sca,LCP}$} (blue line), respectively.}
		At dual behaviour ($\cancel{d}=0$), the chirality conversion $\Xconv$ vanishes, and anti-duality (minimal $\cancel{a}$) is observed as a minimum in $\Xsca$.
	}
	\label{fig:Mie_Si}
\end{figure}

\ifthenelse{\boolean{isSciRep}}{(i)}{\subsubsection*{Achiral Isotropic Scatterer}}
Next, we first analyze the scattering by an achiral non-Rayleigh dipolar particle.
Specifically, we address the behaviour of a spherical
silicon (Si) nanoparticle of radius $r=230\text{nm}$ and refractive index $n=3.5$ in the near-infrared regime
illuminated with a CPL plane wave of positive helicity, i.e.~left circularly polarized, LCP(+).
\change{Since the sphere is an achiral scatterer, illumination with a CPL plane wave of negative helicity, i.e.~right circularly polarized, RCP (-),
yields equal results with regards to energy ($\Wsca^- = \Wsca^+$), while all chiral quantities present an additional minus sign (e.g.~$\Xsca^-=-\Xsca^+$).}

It has been shown that lossless dielectric particles of high  permittivity, like this one, sustain
strong magnetic dipoles and multipoles and hence are suitable constitutive elements for new photonic materials and devices
\cite{evlyukhin2010,garcia2011,kuznetsov2012,decker2016,kuznetsov2016}.
As depicted in Fig.~\ref{fig:Mie_Si}(b), the particle shows dipolar behaviour at incident wavelengths larger than 1200nm,
exhibiting an electric and magnetic dipole peak at 1280nm and 1680nm, respectively \cite[reproduced here as Fig.~\ref{fig:Mie_Si}(b) to ease understanding]{garcia2011}.

Fig.~\ref{fig:Mie_Si}(a) shows that the particle has $\cancel{d}=0$ only at the incident wavelength
$\lambda \approx 1830\text{nm}$, where it is dual. Here, the first Kerker condition is fulfilled since the electric and magnetic
total scattering cross-sections are equal [Fig.~\ref{fig:Mie_Si}(b)] for the incident LCP(+) plane wave.
Since the latter equivalence is necessary but not sufficient for zero-backscattering, the dual behaviour is more
intuitively observed by noting that the incident polarization remains unchanged; i.e.~the scattered field has the same helicity
as the incident wave and, hence, the  chirality conversion due to the scattering is zero, \change{$\Xconv^+=0$} [see Fig.~\ref{fig:Mie_Si}(c)].

Concerning the second Kerker condition, manifested as a minimum of the differential scattering cross-section
in the forward direction, here observed at $\lambda \approx 1550\text{nm}$ where, once again,  the electric and magnetic parts of the scattered energy are equal [see Fig.~\ref{fig:Mie_Si}(b)],
the anti-duality breaking $\cancel{a}$ is at a local minimum of 0.59 [cf.~Fig.~\ref{fig:Mie_Si}(a)]. However, the breaking of helicity annhilation $\cancel{h}_a$ is at a local maximum.
Note in this connection that  a fully anti-dual scatterer yielding $\cancel{a}$=0, would give rise to  $\cancel{h}_a=1$. Nevertheless,  due to causality, a fully anti-dual behaviour
is unphysical for lossless particles \cite{nieto2011}.  A perfect anti-dual scatterer would convert the incident circular polarization fully into its opposite handedness.
Here, it should be stressed that the sign of the scattered chirality \change{$\Xsca^+$} \eqref{eq:dip_Xsca} involves both amplitudes and phases of the polarizabilities.
Accordingly, the scattered chirality is at a minimum for this wavelength of minimum forward scattering [cf.~Fig.~\ref{fig:Mie_Si}(c)].

Furthermore, depolarization effects such as Rayleigh depolarization at long wavelengths \cite{laan2015}, are described by the breaking of helicity annihilation $\cancel{h}_a$.
In Fig.~\ref{fig:Mie_Si}(a), it is observed that at both $\lambda \approx 1330\text{nm}$ and $\lambda \approx 1640\text{nm}$ the scattered light is achiral, i.e.~linearly
polarized. This property is observable as a vanishing scattered chirality $\Xsca = 0$ [Fig.~\ref{fig:Mie_Si}(c)].
At these wavelengths, the contributions of the particle induced  dipoles,  yielding positive and negative helicity in the scattered energy, are equal
and thus cancel each other [Fig.~\ref{fig:Mie_Si}(d)]. As a result, an achiral response of the isotropic dipolar scatterer takes place.

\change{
The interplay between the electric and magnetic induced multipoles in these magnetodielectric particles gives rise to  excitations in the body whose helicity
is either the same or opposite to that of the incident wave;
namely, dipoles and multipoles induced in the sphere,  that represented in the helicity basis [cf.~\eqref{eq:helitrans}] are of positive [LCP(+)] or negative [RCP(-)] helicity.
Both kinds of excitations appear in  the case of incident LCP(+) plane wave illumination. For example, a LCP(+) dipole is due to an electric dipole $\vec{d}$ which oscillates
with a positive phase shift of $\pi$ with respect to a magnetic dipole $\vec{m}$ of equal amplitude ($\vec{d} = i\vec{m}/c$) \cite{fernandez2013,wozniak2018}.
Conversely, a negative helicity  RCP(-)   electric dipole has a phase shift of $-\pi$ compared to the corresponding magnetic dipole, i.e.~$\vec{d} = -i\vec{m}/c$.
Thus we denote their positive and negative helicity property with \textit{subscripts} $\text{LCP}$ and $\text{RCP}$, respectively.
In general, both kind of dipoles are induced by an incident wave of given helicity, and they determine the polarization of the \textit{scattered} field.
We recall that as introduced before, the \textit{superscripts} $\pm$ stand for \textit{incident} light of positive and negative helicity, respectively.
\\ \indent
The dipoles and multipoles of helicity opposite to that of the incident wave are associated to the conversion of chirality (or helicity) on scattering.
This phenomenon is seen when we  compare this chirality conversion \change{$\Xconv^+$}
[cf.~red line in Fig.~\ref{fig:Mie_Si}(c)] and the excitation of \change{induced} dipoles and quadrupoles, which expressed in the  helicity basis have negative handedness,
[see $W^\text{+,dip}_\text{sca,RCP}$ and $W^\text{+,qua}_\text{sca,RCP}$, red and pink lines, respectively, in Fig.~\ref{fig:Mie_Si}(d)].
On the basis of the criterion above, these two latter excitations have helicity [RCP (-)] opposite to that of the incident wave which has been chosen as positive [LCP (+)], as explained before.
}

Hence, taking into account that the scatterer is lossless and that, as just seen, the scattered energy consists of two parts: one of negative and one of positive helicity, 
it follows that the total scattered energy $\Wext^+$ due to this incident plane wave with positive helicity LCP(+) is written as:
\change{$\Wext^+ = \Wsca^+ = W^+_\text{sca,LCP} + W^+_\text{sca,RCP}$}.
Furthermore, the scattered chirality measures the difference of these two contributions \cite{gutsche2016a}: \change{$\Xsca^+ \propto W^+_\text{sca,LCP} - W^+_\text{sca,RCP}$}.
From the equivalence of extinction of both energy and  chirality for incident fields of well-defined helicity [cf. \eqref{eq:extHelicityMatrices}],
we deduce that \textit{for a lossless scatterer the chirality conversion is given by twice the scattered energy of opposite helicity}, i.e.~\change{$\Xconv^+ \propto 2 W^+_\text{sca,RCP}$}.
This is  confirmed in the dipolar regime ($\lambda > 1200\text{nm}$), as well as around $1170\text{nm}$ where the magnetic quadrupole dominates;
and shows the significance of this novel observable, \change{namely the chirality conversion}, specifically for scattering objects that, like these high refractive index dielectric particles,
exhibit a strong magnetic response to the incident field.

\begin{figure}[ht]
	\centering
	\includegraphics[width=.85\textwidth]{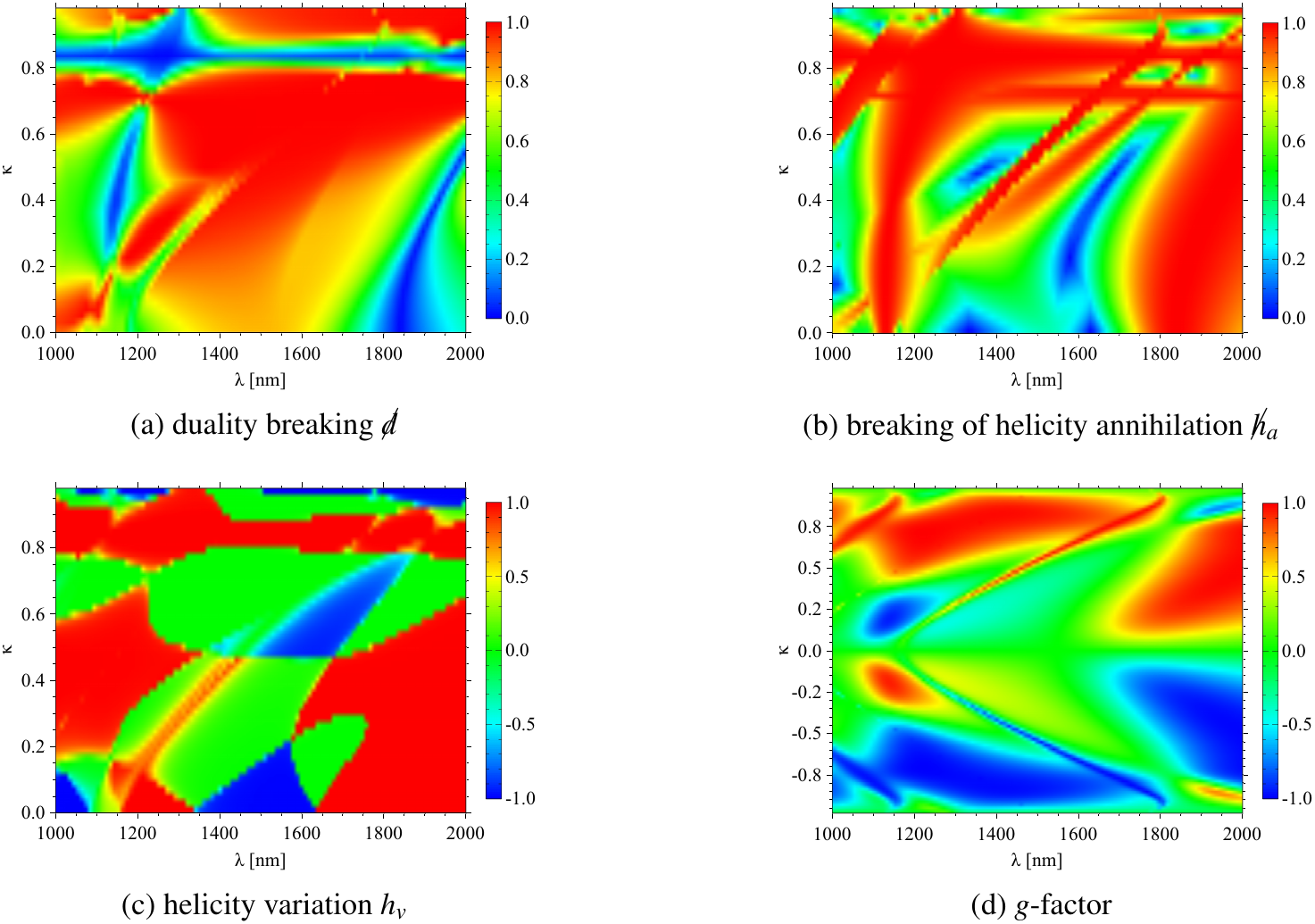}
	\caption{
		Spherical chiral particle of radius $r=230\text{nm}$, refractive index $n=3.5$ and varying  chirality parameter $\kappa\neq 0$.
		Colorbars indicate breakings of:  (a) duality $\cancel{d}$ and (b) helicity annihilation $\cancel{h}_a$;  as well as
		(c)  helicity variation $h_v$ and (d) the $g$-factor. The  wavelength at which  the achiral particle ($\kappa= 0$) behaves as  dual is  red-shifted,
		and for a large  $\kappa$ the particle is dual independently of the wavelength.
		The long wavelength achiral helicity annhiliation is also red-shifted, whereas the short wavelength achiral peak is vanishing for  increasing chirality.
		The helicity variation resembles the qualitative behaviour of duality. The quantities
		$\cancel{d}, \cancel{h}_a$ and $h_v$ are symmetric with respect to $\kappa$;  while
		the $g$-factor changes sign for positive and negative $\kappa$, and, being  strongly wavelength dependent, it generally increases as so does $|\kappa|$.
	}
	\label{fig:Mie_ch}
\end{figure}

\ifthenelse{\boolean{isSciRep}}{(ii)}{\subsubsection*{Chiral Isotropic Scatterer}}
Now, we introduce chirality in the geometrically isotropic scatterer by means of a non-zero chirality  parameter $\kappa$,
while this body remains reciprocal  and bi-isotropic \cite{lindell1994}. For lossless media, the absolute value of $\kappa$ is smaller than unity \cite[Eq.~(D.8)]{lindell1994}, as stated above.
Here,  as before, we fix the sphere radius $r=230\text{nm}$ and its refractive index $n=3.5$. The chirality parameter $\kappa$ is then varied from zero
(which would be the  case discussed above) to unity. Of course now the particle will no longer be made of Si,
since $\kappa \neq 0$ makes the material hypothetical.
Nevertheless, this yields a direct way of studying the interplay of significant magnetodielectric effects with phenomena stemming from the  chirality of the scattering object.

The breakings of duality and helicity annihilation, as well as the helicity variation, are shown in Fig.~\ref{fig:Mie_ch}(a)-(c).
These quantities are symmetric with respect to $\kappa=0$, while
the $g$-factor is antisymmetric [Fig.~\ref{fig:Mie_ch}(d)].
For vanishing $\kappa$, the scatterer is achiral and, accordingly,  $g=0$. The magnetic quadrupole resonance at 1160nm of the achiral particle
is clearly red-shifted and yields very large $g$-factors up to one. Interestingly, the sign of $g$ is not directly coupled to the sign of the chirality parameter,
but changes over the analyzed spectrum: regions of highly positive, as well as of very negative, $g$-factor occur throughout this parameter domain.
In this regard, it should be noted that natural materials usually posess very small effective chirality parameters \cite{schaeferling2016}, these being in the range of $10^{-3}$.

A very unusual behaviour is observed in the duality of this magnetodielectric object with high $\kappa$ [Fig.~\ref{fig:Mie_ch}(a)]: at $\kappa \approx 0.85$,
the scatterer is dual irrespective of the incident wavelength.
\changeTwo{
Due to the red-shifted interference of magnetic quadrupole and electric dipole resonances, the particle is even stably dual with respect to both $\lambda$ and $\kappa$ in this regime.
This stable minimum in $\cancel{d}$ is visible as a cross-like structure centered at $\lambda \approx 1250\text{nm}$ and $\kappa \approx 0.85$.
}
Although in the range of wavelengths  in this study, the particle shows either dipolar or quadrupolar behaviour depending on $\lambda$,
the stability of its dual behaviour is more intuitively understood in the dipolar domain. For $\kappa > 0.85$, the electric polarizability is dominant,
whereas for chirality parameters smaller than 0.85, the magnetic response is larger (not shown for the sake of brevity).
The change from a predominantly electric response to a magnetic one occurs at $\kappa=0.85$ throughout the spectrum.
Here, the first Kerker condition $\alpha_e = \alpha_m$ yields vanishing duality breaking [cf. Eq.\eqref{eq:dip_d}] with cross electric-magnetic polarizibilities $\alpha_{em}$
of an order similar to that of the electric and magnetic ones. Additionally, the achiral dual resonance at 1830nm is again red-shifted.

This overall red-shift is also apparent in the breaking of helicity annihilation [Fig.~\ref{fig:Mie_ch}(b)]. The first helicity
annihilation resonance at $\lambda \approx 1330\text{nm}$
vanishes as the chirality parameter shows up because the red-shifted magnetic quadrupole peak is then larger and exhibits a non-zero scattered helicity.
Concerning the annihilation peak at 1640nm in the achiral case [cf.~Fig. 1(a)], it is stable up to $\kappa\approx0.6$.
For larger chirality parameters, the resonance of helicity annihilation vanishes and barely  there are  regimes of similar behaviour,
i.e.~most scattered light is not linearly polarized for incident fields of well-defined helicity.
The general trends of both duality and helicity annihilation breakings are also visible in the helicity variation, however, since this
condition is weaker for both dual and helicity annihilating behaviour, as discussed before, the areas of extreme $h_v$-values are better distinguishable with sharp transitions
from positive to negative $h_v$ [Fig.~\ref{fig:Mie_ch}(c)].

\ifthenelse{\boolean{isSciRep}}{}{\subsection*{Anisotropic Scatterer}}

The study of non-spherical anisotropic scatterers is highly dependent on the direction of illumination. Chiral molecules are often
analyzed in dilute solutions wherein they are randomly oriented. Mostly, only small numbers of single molecules are available
and multiple scattering can be neglected. The quantities introduced above are suitable for the analysis
of such isolated scatterers, as we show in the next two illustrations: (iii) and (iv).

\change{
We wish to remark that in recent experiments, the helicity components of the scattered light were measured in transmission \cite{tischler2014} and discussed by dual symmetry \cite{fernandez2013a}.
Furthermore, a chirality flux spectroscopy, measuring the third Stokes parameter, was used to analyze the chiroptical response of two-dimensional chiral structures \cite{poulikakos2018}.
We note that  the helicity of light has been measured mostly in transmission. By contrast, our study addresses the importance of the full angular averaged
chirality flux. This quantity  may be measured in an integrating sphere, or by combining measurements in forward and backward directions as in Ref.~\cite{wozniak2018}.
}

\begin{figure}[ht]
	\centering
	\includegraphics[width=.6\textwidth]{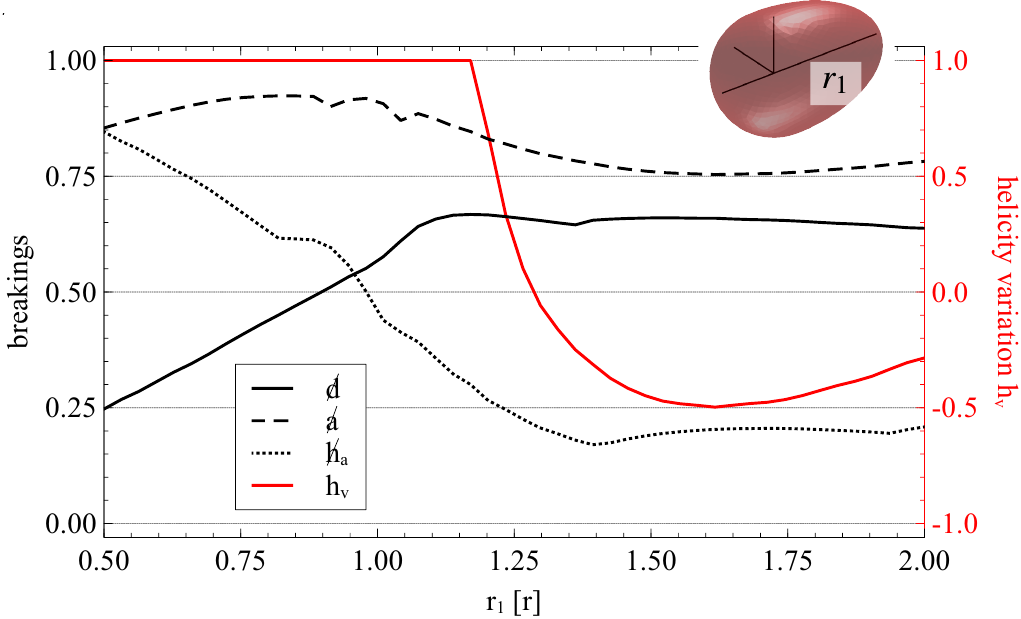}
	\caption{
		Convex body with shrinked ($\frac{r_1}{r}<1$) and elongated ($\frac{r_1}{r}>1$) radius $r_1$ in one direction starting from a sphere with $r=230\text{nm}$ (cf.~main text).
		The incident wavelength is 1680nm and the electric permittivity $\epsi=\sqrt{3.5}$. The breakings of duality $\cancel{d}$, anti-duality $\cancel{a}$ and helicity annhilation $\cancel{h}_a$,
		as well as the helicity variation $h_v$ are displayed. Prolated shapes are helicity-keeping and
		approach dual behaviour. The dominating sign of the scattered chirality is flipped with respect to that of the incident light
		($h_v<0$) for $\frac{r_1}{r}>1.3$. However, this scattered light is depolarized with a breaking of helicity annihilation of approximately 0.2.
	}
	\label{fig:Ani_elli}
\end{figure}

\ifthenelse{\boolean{isSciRep}}{(iii)}{\subsubsection*{Achiral Anisotropic Scatterer}}
Next, we first numerically study the scattering from a convex body with the help of the Finite Element Method (cf.~section Methods).
The six different radii of a generalized ellipsoid are fixed except  the one: $r_1$. Again, we start from the magnetodielectric sphere with $r=230\text{nm}$
and refractive index $n=3.5$. As this object is transformed into an ellipsoid, the varying radius $r_1$ is shrinked ($\frac{r_1}{r} < 1$) to its half and stretched ($\frac{r_1}{r} > 1$)
to its double value. For simplicity,
we fix the incident wavelength to 1680nm, which, as seen above, coincides with the resonance peak of the magnetic dipole when this object is \changeTwo{spherical.}

Since there exist at least two mirror-planes in this geometrically achiral body, both the chirality and $g$-factor vanish.
As shown in Fig.~\ref{fig:Ani_elli}, the scatterer is perfectly helicity-keeping with $h_v=1$ for $\frac{r_1}{r} < 1.2$.
The duality breaking decreases for prolate shapes with a minimal value of 0.2 at half the initial sphere radius. Above $\frac{r_1}{r}=1.2$,
the duality breaking remains nearly constant, but the helicity variation reveals that the incident helicity sign changes in the scattered field for $\frac{r_1}{r} > 1.3$.
However, the object is not anti-dual in the strict sense, since the anti-duality breaking is larger than $0.8$ in the full parameter space, and only the dominating sign
of the eigenvalues of the scattered chirality is represented in the helicity variation
Furthermore, there is no helicity annihilation at this wavelength and geometric variations. The minimal value of the breaking of helicity annihilation is approximately 0.2 for elongated shapes.

\ifthenelse{\boolean{isSciRep}}{(iv)}{\subsubsection*{Chiral Dual Anisotropic Scatterer}}
In what follows we illustrate our formalism for chiral anisotropic particles. We start by analyzing their duality breaking.
A chiral gold particle resulting from a sophisticated fabrication procedure, and studied experimentally \cite{mcpeak2014},  is made dual by introducing
a magnetic permeability. Namely, the particle is analyzed for an incident wavelength of 600nm,
the permittivity $\epsi$ is kept constant at its value for gold at 600nm,  while the permeability varies as $\mue=x(\epsi -1) +1$.
The parameter $x$ is a parameter that grows between 0 and 1, so that when $x=1$, the particle is dual since $\epsi=\mue$ and thus $\alpha_e=\alpha_m$
\cite{kerker1983}.

\begin{figure}[ht!]
	\centering
	\includegraphics[width=.95\textwidth]{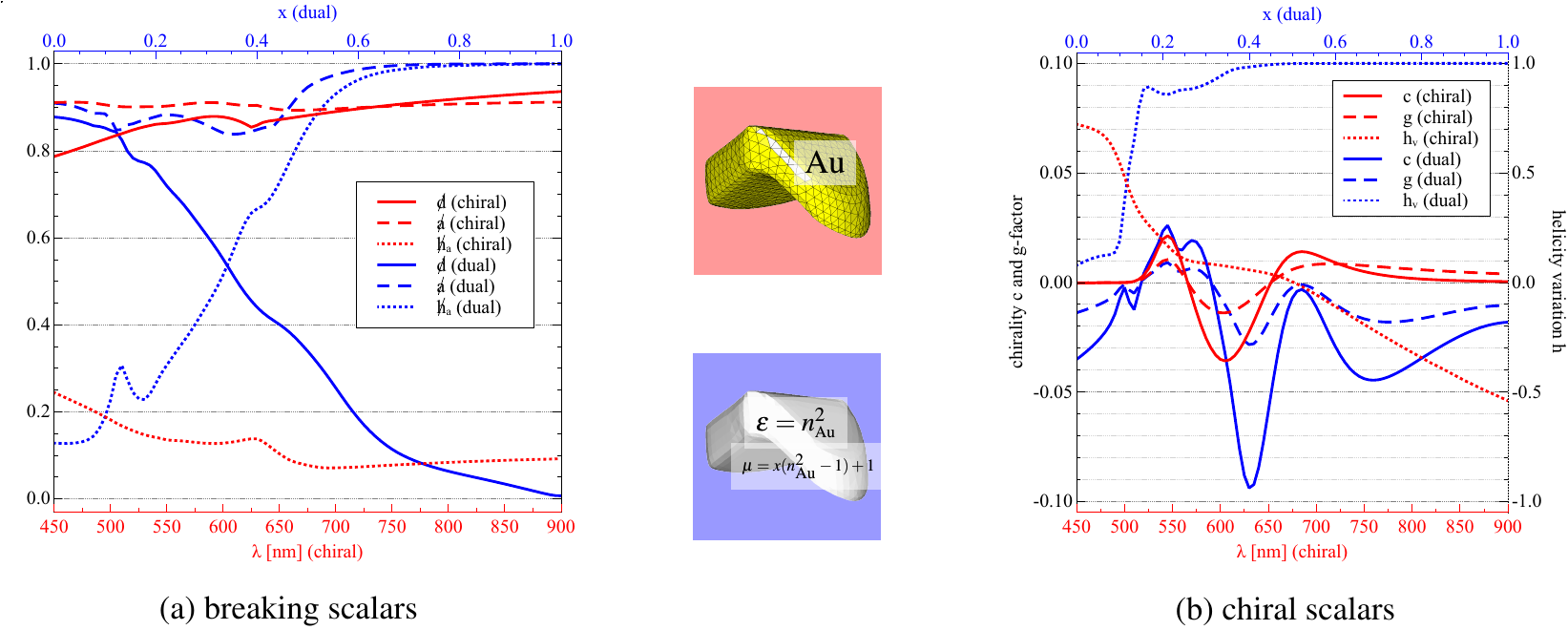}
	\caption{
		Geometrically chiral pyramidal  particle made from gold  \cite{mcpeak2014} (upper inset and red lines) or from a dual material obtained by introducing
		a magnetic permeability at $\lambda=600\text{nm}$ (lower inset and blue lines).
		(a) Breakings of duality $\cancel{d}$, anti-duality $\cancel{a}$ and helicity annihilation $\cancel{h}_a$ dependent both on the
		wavelength for the gold particle and on the duality parameter $x \in [0,1]$ for the dual particle, respectively. (b) Chirality $c$,
		$g$-factor (left axis), and helicity variation $h$ (right axis). The gold particle is neither dual ($\cancel{d}\neq0$)
		nor anti-dual ($\cancel{a}\neq0$)	but scatters mostly achiral light since $\cancel{h}_a \approx 0.15$ over the whole sprectrum.
		Its optically chiral response has been measured experimentally \cite{mcpeak2014} and fits the simulated data.
		The dual particle illustrates the simultaneous conditions $\cancel{d}=0, \cancel{a}=1$ and $h=1$ for a dual symmetric scatterer.
		Throughout the transition from gold ($x=0$) to a dual particle ($x=1$), the particle is dominantly negative chiral.
		The increase of the magnitude of chirality by a factor of six is partly due to the higher refractive index.
	}
	\label{fig:Ani_ptcl}
\end{figure}

As shown in Fig.~\ref{fig:Ani_ptcl}(a) (blue lines), the duality breaking accurately depicts both the non-dual and dual behaviour of the particle as $x$ varies from zero up to one ($\cancel{d}=0$).
The helicity variation with $h_v=1$ is, as discussed before, a weaker condition for duality.
\change{The scatterer is already perfectly helicity-keeping for $x$ larger than 0.5 [see Fig.~\ref{fig:Ani_ptcl}(b)].}
From this value of $x$ on, more detailed information is contained in $\cancel{d}$. However, we can
deduce from the spectrum of helicity variation that the scattered field is dominantly positive chiral for all artificial materials analyzed in this example, since $h_v>0$
for all $x$. Furthermore, the anti-duality and helicity annihilation breakings are unity for $x>0.7$, as expected for dual symmetric materials.

The chirality $c$ and the $g$-factor reveal the highly chiral response of the particle. Large magnitudes of $g$-factors up to 0.1 are reached [see Fig.~\ref{fig:Ani_ptcl}(b)].
This is partly
attributed to the high scattering due to the very large refractive index $n \approx -10.14 + 1.38i$.
Nevertheless, record $g$-factors with values up to 0.02 are also attained when the particle is purely made of gold. This $g$ derived from the full $T$-matrix [cf. Eq. \eqref{eq:g}]
can be interpreted as the average differential response due to incident right and left circularly polarized plane waves.
Our simulated data of $c$ and $g$ correspond well with the
experimentally measured results \cite[cf. Figs.~3(b),(d)]{mcpeak2014}.

\ifthenelse{\boolean{isSciRep}}{}{\subsubsection*{Chiral Anisotropic Scatterer}}

The helicity variation for the chiral gold particle shows a monotonic decreasing behaviour with a sign change at 680nm. For smaller wavelengths the scattered
chirality flux mostly has positive helicity, whereas in the longer wavelength regime a negative helicity is observed. This yields near-fields which are dominated by
positive and negative optical chirality hot-spots, respectively. Note, however, that this is only the case for the scattered field. The coupling of chiral molecules in the near-field
is due to the total field composed of the incident and the scattered one \cite{tang2010}. Accordingly, interference effects might result in helicity-flipping fields
for small wavelengths, even though $h_v>0$. Nevertheless, the study of the helicity variation provides further insight than the study of the previously introduced breakings such
as $\cancel{d}$ in this example. The breakings are constant over the spectrum of incident wavelengths and are not suitable for classifying the scattering response with respect to
polarization phenomena, [see Fig.~\ref{fig:Ani_ptcl}(a), red lines].

In conclusion, we have put forward the quantities derived from the conservation of optical chirality (or helicity) of monochromatic wavefields, both in Mie and $T$-matrix theories.
Such quantities are scalars that characterize the response of arbitrary scattering particles. In particular, the breaking of
dual symmetry is accompanied by a weaker condition based on the scattered chirality, namely, the so-called helicity variation.

Directional effects of the differential scattering cross-section, such as the Kerker conditions, have been included in this general formalism,
both for  achiral and chiral magnetodielectric dipolar spherical particles, and discussed with respect to the experimentally accessible data of scattered  and extinguished energy and chirality.
This has been  illustrated   with high permittivity, non-Rayleigh, dipolar spheres, which have generated so much interest in nanophotonics.
We have shown how their illumination with chiral light  uncovers new important phenomena associated to their  chiral dipolar and quadrupolar resonant excitations,
either  with the same or with  opposite helicity with respect to that of the illuminating wave. 

Anisotropic scatterers, either with or without a mirror-plane, are classified by these novel quantities, and the implications for the scattered fields are discussed.
As an illustration, a comparison with experimental data from a chiral gold particle has been drawn.
We expect that this general framework enables the characterization of a broad range of scattering objects with applications
in chiral molecular spectroscopy, spin photonics, and the design of optical sources as well as of metamaterials and other composite media.

\section*{Methods}

\subsection*{Isotropic Mie Code}

As described in the main text, the submatrices of the $T$-matrix of a geometrically isotropic scatterer are diagonal.
The coefficients are given by \cite[p.~188, adopted to non-even/-odd basis $\vec{M},\vec{N}$]{bohren1940}
\begin{align}
  \alpha_n = \frac{V_n(m_R) A_n(m_L) + V_n(m_L) A_n(m_R)}{W_n(m_L) V_n(m_R) + V_n(m_L) W_n(m_R)} \\ 
  \beta_n = \frac{W_n(m_L) B_n(m_R) + W_n(m_R) B_n(m_L)}{W_n(m_L) V_n(m_R) + V_n(m_L) W_n(m_R)} \\ 
  \gamma_n = \frac{W_n(m_R) A_n(m_L) - W_n(m_L) A_n(m_R)}{W_n(m_L) V_n(m_R) + V_n(m_L) W_n(m_R)}.
\end{align}
The coefficients $\alpha_n, \beta_n$ and $\gamma_n$ are the main diagonal elements of $T_{ee}, T_{mm}$ and $T_{em}$, respectively.
The relative refractive indices $m_L$ and $m_R$ and the mean refractive index $m$ are then defined as
$m_L = \frac{n_L}{n_s},
m_R = \frac{n_R}{n_s}$
and
$\frac{1}{m} = \frac{1}{2} \left( \frac{1}{m_R} + \frac{1}{m_L} \right) \frac{\mue}{\mue_s}$,
where $n_s$ and $\mue_s$ are the refractive index and the relative
permeability of the surrounding medium, respectively. The results in Fig.~\ref{fig:Mie_Si} and \ref{fig:Mie_ch} are obtained
using these analytic solutions of Maxwell's equations.

\subsection*{Anisotropic Numerical Simulations}
The solution of time-harmonic Maxwell's equations for arbitrary geometries are obtained with the commerical solver
\textit{JCMsuite} based on the Finite Element Method (FEM). The particles analyzed in Fig.~\ref{fig:Ani_elli} and \ref{fig:Ani_ptcl} are
discretized by tetrahedral meshes. Convergence is ensured for sidelength constraints of $h=50\text{nm}$ and
ansatz functions of polynomial degrees $p=3$ and $p=4$ for Fig.~\ref{fig:Ani_elli} and \ref{fig:Ani_ptcl}, respectively. The open boundary conditions are modelled by
Perfectly Matched Layers which are controlled by a software-specific precision parameter\cite{JCMsuite} of $1e-5$.

The $T$-matrices are computed by  illumination with 36 and 88 plane waves, respectively. These are equally distributed on a sphere in
$k$-space, and are created with random polarizations. This procedure yields accurate results up to multipoles of order $m=3$
for the achiral ellipsoid and $m=5$ for the chiral particle, respectively. The entries of the $T$-matrix are efficiently
calculated by a surface integral based on the conservation of extinction \cite{garcia2018}.

\ifthenelse{\boolean{isSciRep}}{}{\bibliographystyle{abbrv}}

\section*{Acknowledgements}

We acknowledge support by Freie Universit\"at Berlin through the Dahlem Research School and 
by MINECO-FEDER, grants \changeTwo{FIS2014-55563-REDC}, and FIS2015-69295-C3-1-P.
We thank Sven Burger for fruitful discussions and Xavier Garcia Santiago for his work on the Mie code and the algorithm for obtaining $T$-matrices from FEM simulations.

\ifthenelse{\boolean{isSciRep}}{
	\section*{Author contributions statement}
	P.G.~and M.N.-V.~conceived the idea of this study. P.G.~developed the $T$-matrix formalism and performed the simulations. P.G.~and M.N.-V.~analysed the results and reviewed the manuscript. 

	\section*{Additional information}
	The authors declare that they have no competing interests.
}{}

\end{document}